\documentclass[twocolumn,showpacs,preprintnumbers,amsmath,amssymb,superscriptaddress]{revtex4}

\usepackage{graphicx}
\usepackage{dcolumn}
\usepackage{bm}

\begin{document}

\preprint{APS/}

\title{Nonlinear Optical Response of SrTiO$_{3}$/LaAlO$_{3}$ Superlattices}

\author{N.~Ogawa}
 \email[]{ogawa@myn.rcast.u-tokyo.ac.jp}
\author{K.~Miyano}
\affiliation{Research Center for Advanced Science and Technology (RCAST), 
University of Tokyo, Komaba, Meguro-ku, Tokyo 153-8904, Japan}
\affiliation{CREST, Japan Science and Technology Agency,
4-1-8 Honcho, Kawaguchi, Saitama 332-0012, Japan}

\author{M.~Hosoda}
\author{T.~Higuchi}
\affiliation{Department of Advanced Materials Science, 
University of Tokyo, Kashiwa, Chiba 277-8651, Japan}

\author{C.~Bell}
\author{Y.~Hikita}
\author{H.~Y.~Hwang}
\affiliation{CREST, Japan Science and Technology Agency,
4-1-8 Honcho, Kawaguchi, Saitama 332-0012, Japan}
\affiliation{Department of Advanced Materials Science, 
University of Tokyo, Kashiwa, Chiba 277-8651, Japan}

\date{\today}

\begin{abstract}
The electronic symmetry of the SrTiO$_{3}$/LaAlO$_{3}$ interface was investigated by optical second harmonic generation, using superlattices with varying periodicity to study the evolution of the electronic reconstruction while avoiding substrate contributions. The superlattices show large perpendicular optical nonlinearity, which abruptly increases when the sublattice thickness goes above 3 unit cells, revealing substantial effects of the polar-nonpolar interface. The nonlinear 'active' area is primarily in SrTiO$_{3}$, develops with increasing thickness, and extends up to 8 unit cells from the interface.
\end{abstract}

\pacs{73.20.-r, 78.20.-e, 42.65.Ky}
\maketitle
Although well developed in conventional semiconductors, engineering of oxide interfaces through the control of atomic termination has opened a new strategy to create novel electronic phases confined at the nanoscale. As an example, the emergence of an electron gas~\cite{AOhtomo} which is superconducting~\cite{NReyren, ADCaviglia} at the interface between two band insulators SrTiO$_{3}$~(STO) and LaAlO$_{3}$~(LAO) has been a subject of intense debate. The contributions from the polar discontinuity, oxygen vacancies, interdiffusion, and concomitant lattice distortions were studied both experimentally~\cite{NNakagawa, JLMaurice, SThiel, MHuijben, ABrinkman, GHerranz, WSiemons, AKalabukhov, PRWillmott, MBasletic, KYoshimatsu, VVonk} and theoretically~\cite{RPentcheva, JMAlbina, MSPark, Ishibashi, ZSPopovic, JLee}. Two types of heteropolar interfaces can be prepared in this system: (TiO$_{2}$)$^{0}$/(LaO)$^{+}$ ($n$-type) and (AlO$_{2}$)$^{-}$/(SrO)$^{0}$ ($p$-type), where $\mp$0.5$e$ charge is nominally expected to be transferred at the interfaces to avoid the potential divergence~\cite{NNakagawa}. However, only the $n$-type interface was found to be conducting. In reality, some lattice polarization should be involved to screen the potential difference, thus giving a complex electronic and structural reconstruction as predicted theoretically~\cite{MSPark, Ishibashi, ZSPopovic, JLee, SOkamoto, DRHamann}. For this interface, recently it was shown that the metallic state is confined within a few nanometers for oxidized samples~\cite{MBasletic}, confirming its 2-dimensional character. Although experimental and theoretical results are beginning to converge, many open questions on the nature of this interface remain.  

One clue to address this issue is the ``critical thickness effect''; the metallic state only emerges for LAO films on STO substrates with thickness more than 3 unit cells~\cite{SThiel}, and a similar but gradual threshold was reported for the case of complementary interfaces~\cite{MHuijben}. This criticality involves both electronic and lattice instabilities, and probing this transition can provide fundamental insight on the metallic state. Optical second harmonic generation (SHG) is a versatile and nondestructive probe to study buried interfaces with atomic-layer sensitivity~\cite{Shen}. Since SHG is forbidden in a centrosymmetric material in the electric dipole approximation, electronic symmetries of surfaces or interfaces of such materials, even with free carriers, can be selectively detected. It has also been utilized for the interfaces of correlated oxides~\cite{HYamada}, with a scale down to a single interface~\cite{NOgawa}. Thus SHG is a powerful probe of the electronic structure of the LAO/STO interface~\cite{ASavoia}.

When probed with nonlinear optics, great care is needed for STO. Bulk STO is paraelectric at all temperatures due to quantum fluctuations~\cite{KAMuller}. However, with compressive or tensile strain~\cite{JHHaeni, AVasudevarao, DGSchlom}, under external electric field~\cite{TSakudo}, or on polarizable substrates~\cite{TZhao}, STO undergoes a ferroelectric transition at finite temperatures. In addition, any defects in STO have substantial contributions to SHG~\cite{WPElffroth}. In practice, a single $n$-type interface shows relatively large SHG, which highly depends on the substrates and annealing conditions as will be discussed later. To avoid this large variable contribution from STO substrates, we have employed superlattice structures, in which the SH coherence length ($\sim$ 40 nm)~\cite{EDMishina2} is within the total thickness of the SLs ($>$ 55 nm) [Fig.~1(a)], thus removing the substrate contribution from the signal.

\begin{figure}
\includegraphics[scale=1]{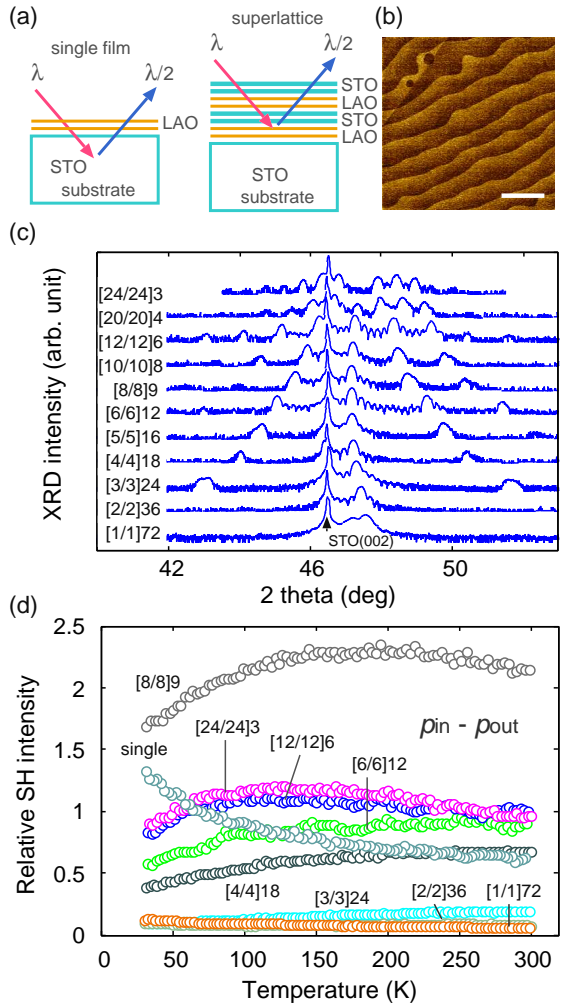}
\caption{\label{fig1}(color online) (a) Schematic illustrations of second harmonic generation from a thin single film and a superlattice. (b) A representative topographic AFM images for the [STO(6)/LAO(6)]12 sample. A scale bar of 1 $\mu m$ is indicated with a white line. (c) Superlattice profiles in the x-ray diffraction. (d) Temperature dependence of the SH intensity in $p_{in}$-$p_{out}$ geometry for SLs with total thickness of 144 unit cells [series (i)], together with that for the single $n$-type interface.}
\end{figure}

Several series of SrTiO$_{3}$/LaAlO$_{3}$ SLs were grown on TiO$_{2}$-terminated STO(001) substrates by pulsed laser deposition with oxygen pressure of $1.0\times 10^{-5}$ Torr and substrate temperature at 973-1023 K. The thickness of each layer was controlled by monitoring the reflection high-energy electron diffraction (RHEED) intensity oscillations, which was confirmed by x-ray diffraction measurements [Fig.~1(c)]. The fabricated SLs are denoted by [STO($k$)/LAO($l$)]$m$, or simply [$k$/$l$]$m$, where $k$ and $l$ refer to the thickness in unit cells (uc), and $m$ indicates the number of periods. We show the results for three series of SLs: (i) total thickness of 144 uc with the ratio of STO:LAO = 1:1, i.e., [1/1]72, [2/2]36 to [24/24]3, (ii) total thickness of 160 uc with STO:LAO = 1:1, and (iii) with different ratios of STO:LAO ([6/12]8 and [12/6]8). The motivation for series (i) and (ii) was to keep the total thickness of thin film STO, LAO, and the growth conditions the same. This allows us to keep growth kinetic effects such as possible oxygen vacancies constant, and quantitatively compare the optical response as a function of interface separation. Single interface $n$-type and $p$-type samples, LAO(16 uc)/STO and LAO(16 uc)/SrO/STO, were also prepared for reference. Transport measurements of the SLs show a sudden increase in sheet resistance around 3 uc, which is in agreement with previous reports~\cite{SThiel, MHuijben}. All fabricated samples were carefully checked with atomic force microscopy (AFM), and we confirmed that their surfaces are terminated with clear step-terrace structures with step height of $\sim$0.4 nm [Fig.~1(b)]. Since the surface quality is similar for all films, we can attribute the difference in nonlinear optical signals to the internal or interface electronic states of the SLs.

These samples were mounted in an ultrahigh vacuum cryostat, and SHG was measured with 1.55 eV fundamental light (150 fs duration at 1 kHz repetition rate) incident through a $\lambda$/2 plate and a lens with 90-degree reflection geometry (0.5$\sim$1.5 mW on $\sim$80 $\mu$m spot). The generated SH was directed to a glan prism, color filters, and a monochromator, and detected with a photomultiplier tube. The signal was normalized by that of a reference KDP crystal, and accumulated more than 10$^{4}$ times at each polarization configuration. The SH energy is within the band gaps of STO (3.2 eV) and LAO (5.6 eV). We note that the SHG from a LAO film on (LaAlO$_{3}$)$_{0.3}$(SrAl$_{0.5}$Ta$_{0.5}$O$_{3}$)$_{0.7}$(LSAT)(001) substrate was negligibly small. Commercial STO(001) substrates also typically show smaller SHG compared to those of the SLs, with relatively large sample to sample dependence, likely due to the thermal history and residual stress.  

\begin{figure}
\includegraphics[scale=1]{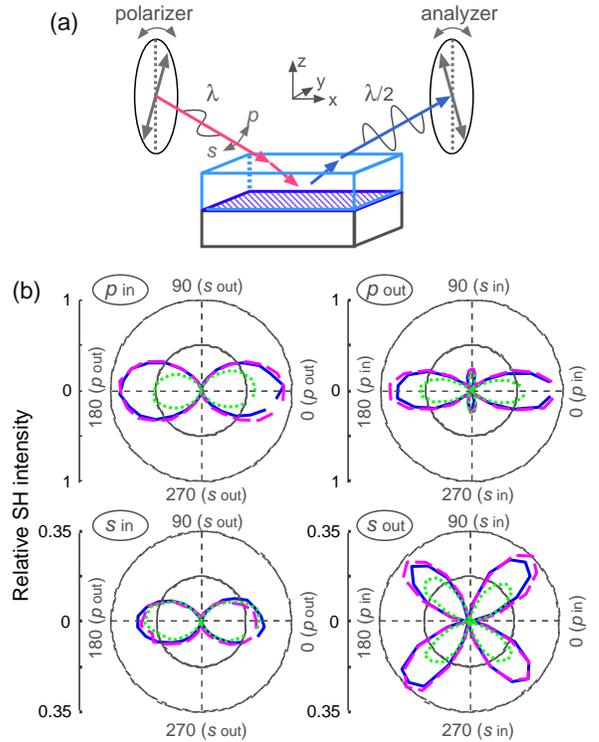}
\caption{\label{fig2}(color online) (a) A schematic of the optical geometry and coordinates for the SH polarization analysis. (b) Polar plots of the relative SH intensity for [STO(24)/LAO(24)]3 (dashed lines), [STO(12)/LAO(12)]6 (solid lines), and [STO(6)/LAO(6)]12 (dotted lines) samples at 30 K.}
\end{figure}

 At the interface of SLs, where inversion symmetry is broken, or under tetragonal distortion, STO has three independent elements in the nonlinear susceptibility tensor, $\chi_{xzx}=\chi_{yzy}$, $\chi_{zxx}=\chi_{zyy}$, and $\chi_{zzz}$ ($4mm$ symmetry). The SH intensity is roughly proportional to $\chi^{2}$, and $p_{in}$-$p_{out}$ optical geometry contains SHG from all these elements.

Figure 1(d) shows the temperature dependence of the relative SH intensity from SL series (i) and a single $n$-type interface with $p_{in}$-$p_{out}$ geometry. Most of the SLs, except for [1/1]72 and [2/2]36, show a decrease in SH intensity at low temperature, while the single $n$-type interface shows relatively large signal, which increases at low temperature. The SH intensity decreases with decreasing thickness for the sublattice thinner than [8/8].

Since the fundamental light penetrates deep in the STO substrate for the case of single interfaces, the increase in the SH intensity at low temperature can have contributions from defects in the substrate~\cite{WPElffroth}. We note that similar behavior can be observed in Nb-doped STO substrates. To confirm this, we post-annealed the single $n$-type sample in 1 atm of oxygen at 773 K for 10 min, and found that the SH intensity decreased by about 50\% at 300 K, and became almost temperature independent (not shown). The single $p$-type interface as grown has $\sim$6 times smaller SH intensity compared to that of the single $n$-type interface at 300 K, and showed little change with temperature and with oxygen annealing. This difference can be due to the fundamental asymmetry of the $n$- and $p$-type interfaces~\cite{NNakagawa}. However, it is difficult to exclude the bulk substrate contributions in single interfaces, and to isolate whether annealing has changed the bulk state, filled oxygen vacancies, or otherwise affected the interface (diffusion or relaxation). We will therefore concentrate on the SL samples hereafter, in which $n$-type interfaces and their spacing dominate the evolution of the SHG, we are free from substrate contributions, and any growth effects are constant.  

Since STO is easily polarized by external perturbation, the likely source of SHG from SLs will be the structural reconstructions coupled strongly with electronic reconstructions at the interface. Experiments~\cite{JLMaurice, PRWillmott, MSalluzzo} and theoretical calculations~\cite{MSPark} reported that the unit cell at the interface is elongated, perhaps due to the Jahn-Teller effect or simply the larger ionic radius of Ti$^{3+}$. The consequence is that the lattice polarization occurs mostly in the STO and that the displacements of the central Ti atom extends several uc from the interface~\cite{ZSPopovic}.

With polarization analysis of the generated SH field, we can extract information about the tensor elements. A schematic of the optical geometry with the experimental coordinates and resultant polar plots of the SH intensities are shown in Figs.~2(a) and (b), respectively, for representative samples. It is clearly seen that all SLs show $4mm$ symmetry at the interface, as expected, just with differences in the amplitude of the tensor elements. We found that, for each SL, $\chi_{zzz}$ is about 40 times larger than the other independent elements, which indicates a large asymmetry normal to the interface. We note that these three elements have similar amplitudes for bulk BaTiO$_{3}$~\cite{Shen} and STO under external electric field~\cite{TSakudo}.

In Fig.~3, we plot the extracted $\chi_{zzz}$ normalized by the total thickness of STO in each SL. First of all, we note that the normalized $\chi_{zzz}$ for [6/12]8 and [12/6]8 (filled marks) have almost the same amplitude, suggesting that the SH is generated in the STO layer with bulk-like contributions. This is consistent with the fact that LAO is far less polarizable than STO. At 30 K, the normalized $\chi_{zzz}$ shows a sudden increase for the sublattice thickness between [3/3] and [4/4] uc, which is reminiscent of the ``critical thickness'' reported~\cite{SThiel,MHuijben}. The normalized $\chi_{zzz}$ shows a peak at the sublattice thickness of 8 uc. Other tensor elements, $\chi_{xzx}$ and $\chi_{zxx}$, show similar trends, indicative of a single origin~\cite{BFLevine}.

\begin{figure}
\includegraphics[scale=1]{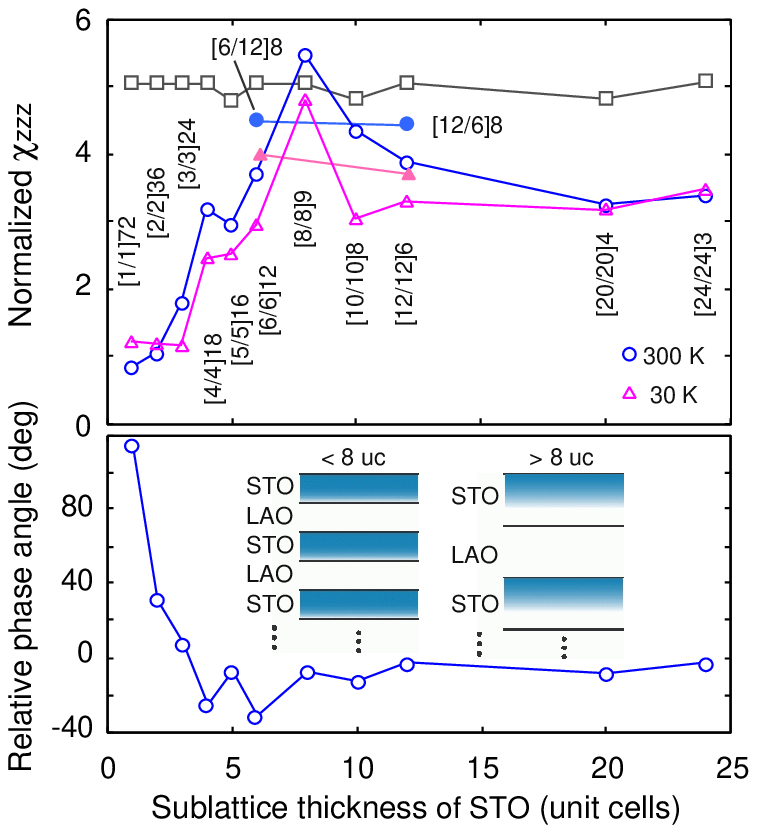}
\caption{\label{fig3}(color online) (Top) Sublattice thickness dependence of the nonlinear susceptibility $\chi_{zzz}$ normalized by the total thickness of STO. Circles and triangles show the data at 300 K and 30 K, respectively, for the SL series (i) and (ii). Filled marks are the data for SL series (iii). Open squares show the results of numerical calculations assuming uniformly polarized STO, scaled appropriately. (Bottom) Relative phase of the SH field in $p_{in}$-$p_{out}$ geometry. The insets show illustrations of the nonlinear layer in SLs with different sublattice thickness.}
\end{figure}

To quantitatively resolve the origin of this structure dependence, we performed numerical calculations for the SHG from STO/LAO SLs~\cite{Bloembergen}. Assuming that STO has a spatially uniform nonlinear susceptibility, and by using realistic parameters, i.e., the experimentally extracted $c$-axis length, 0.3905 nm and 0.375 nm for STO and LAO (strained on STO), respectively, reported dielectric constants, and with multiple reflection and interference rigorously taken into account, we found that the resultant reflection SH intensities are almost independent of the SL structures for the sublattice thicknesses we employed (Fig.~3, top, open squares)~\cite{note1}. This shows clear discrepancy with the experiments. Thus, in our samples, the optical nonlinearity develops with thickness from [1/1]72 to [8/8]9 at 300 K, and from [4/4]18 with a clear threshold at 30 K.  

We also measured the relative phase of the SH field to probe the development of the electronic asymmetry at the interface (Fig.~3, bottom). For this, we utilized an interference technique~\cite{NOgawa, ASavoia} at 300 K. The phase angle rotated continuously from [1/1]72 to [4/4]18 and then saturated. This again shows a clear threshold at the sublattice thickness between [3/3] and [4/4] uc. One possible explanation of this phase change is the phase shift of the SH field through the appearance of free carriers, although the situation is complicated by the simultaneous development of the lattice polarization.   

The salient features of our observations are (1) the development of optical nonlinearity with sublattice thickness, (2) emergence of a critical thickness in the susceptibility and in the SH phase, and (3) the peak structure in the SH intensity, or $\chi_{zzz}$, for the sublattice thickness of [8/8] uc. They can be explained as follows: (a) SHG is primarily from STO under electronic asymmetry, which increases as the interface is approached~\cite{PRWillmott} to screen the polar discontinuity, (b) for thinner sublattices, the asymmetry decreases because of insufficient polar energy to cause the reconstruction, with the effects of complementary interfaces close to each other~\cite{SThiel, MHuijben}, (c) a transition occurs at the critical thickness, which induces lattice polarization and its screening by the free carriers, and (d) the electronic asymmetry extends up to 8 uc in STO, and the apparent nonlinearity decreases for thicker films because the total volume of the nonlinear region decreases (Fig.~3, insets). 

Recently it was reported that the splitting and shift of Ti $d$ levels, discussed as an orbital reconstruction, has a precursor already with 2 uc of LAO on a STO substrate~\cite{MSalluzzo}, which is nearly temperature independent even with thicker coverage. This is consistent with our observations in that the structural and electronic distortions are already triggered within the critical thickness, and that the SH intensity does not show strong temperature dependence. 

In summary, we have studied the electronic asymmetry at the interface of LAO/STO superlattices, and found that the interfaces have large optical nonlinearity with a clear threshold between 3 and 4 unit cells of STO. This criticality is induced by the polar discontinuity and can be understood mainly as a lattice polarization under the influence of induced free carriers.     

\begin{acknowledgments}
We are grateful to T. Susaki and S. Tsuda for preparing single interface samples. This work was partly supported by MEXT TOKUTEI (16076207) and GCOE for Phys. Sci. Frontier, MEXT, Japan.
\end{acknowledgments}

\newpage 

\begin{thebibliography}{99}
\bibitem{note1} The small modulations in the calculated $\chi_{zzz}$ are due to the difference in the total thickness of SLs, which are well reproduced in the experiment.
\end{thebibliography}

\end{document}